\documentclass[aps,prl,twocolumn,,superscriptaddress,floatfix]{revtex4-1}
\usepackage[svgnames]{xcolor}
\usepackage[colorlinks=true,urlcolor = purple,linkcolor=teal,citecolor=magenta,bookmarks=false]{hyperref}
\usepackage{amsfonts,amsmath,amssymb}
\usepackage{graphicx}
\usepackage{dsfont}

\usepackage{lipsum}
\usepackage{graphicx}
\usepackage{amsmath,amssymb}
\usepackage{braket}
\usepackage{mathtools}
\usepackage{hyperref}
\usepackage{multibib}

\usepackage{booktabs,amsmath}

\usepackage{color}
\usepackage[normalem]{ulem}
\usepackage{amsfonts}

\usepackage{pdfpages} % include appendix
\makeatletter
\AtBeginDocument{\let\LS@rot\@undefined}
\makeatother

\definecolor{mypink1}{rgb}{0.858, 0.188, 0.478}
\definecolor{dartmouthgreen}{rgb}{0.05, 0.5, 0.06}
\hypersetup{
    colorlinks=true,
    linkcolor=blue,
    filecolor=blue,      
    urlcolor=mypink1,
    citecolor=dartmouthgreen,
}

\usepackage{pdfpages} % include appendix
\makeatletter
\AtBeginDocument{\let\LS@rot\@undefined}
\makeatother

\begin{document}

\title{Superdiffusion from emergent classical solitons in quantum {spin} chains}

\author{Jacopo De Nardis}
\affiliation{Department of Physics and Astronomy, University of Ghent, Krijgslaan 281, 9000 Gent, Belgium}
\author{Sarang Gopalakrishnan}
\affiliation{Department of Physics and Astronomy, CUNY College of Staten Island, Staten Island, NY 10314;  Physics Program and Initiative for the Theoretical Sciences, The Graduate Center, CUNY, New York, NY 10016, USA }
\author{Enej Ilievski}
\affiliation{Faculty for Mathematics and Physics, University of Ljubljana, Jadranska ulica 19, 1000 Ljubljana, Slovenia}
\author{Romain Vasseur}
\affiliation{Department of Physics, University of Massachusetts, Amherst, Massachusetts 01003, USA }

\begin{abstract}

Finite-temperature spin transport in the quantum Heisenberg spin chain is known to be superdiffusive, and has been conjectured to lie in the Kardar-Parisi-Zhang (KPZ) universality class. Using a kinetic theory of transport, we compute the KPZ coupling strength for the Heisenberg chain as a function of temperature, directly from microscopics; the results agree well with density-matrix renormalization group simulations. We establish a rigorous quantum-classical correspondence between the ``giant quasiparticles'' that govern superdiffusion and solitons in the classical continuous Landau-Lifshitz ferromagnet. {We conclude that KPZ universality has the same origin in classical and quantum integrable isotropic magnets:  a finite-temperature gas of low-energy classical solitons.}

\end{abstract}

\maketitle
\textbf{\textit{Introduction}}.
The dynamics of isolated many-body systems exhibits a remarkable diversity, which we have only begun to understand in the past decade~\cite{pssv, rigolreview, bertini2020finite}. Dynamics in one dimension is particularly rich, as experimental and theoretical studies have shown. Although experiments often deal with systems far from equilibrium~\cite{kinoshita, Hofferberth:2007aa, gring, bloch2014, schreiber2015observation, bernien2017probing, erne2018, tang2018, wilson2019observation, kao2020creating}, from a theoretical perspective it is most natural to characterize dynamics in the linear regime about equilibrium states. Linear response can be probed via transport experiments~\cite{krinner2015observation, hess2019heat} or by measuring dynamical correlations~\cite{zwierlein2019}. Generically, the densities of conserved quantities in lattice models undergo diffusion, as
predicted by linearized hydrodynamics~\cite{PhysRevA.89.053608}. Integrable and many-body localized systems, however, have infinitely many local conserved charges, so simple hydrodynamic arguments fail. Transport is absent in localized systems~\cite{BAA, 2014arXiv1404.0686N, 1742-5468-2016-6-064010, RevModPhys.91.021001} and in general ballistic in integrable systems~\cite{bertini2020finite}, though anomalous transport, including both {\textit{subdiffusion}}~\cite{PhysRevLett.114.100601, Agarwal, gopalakrishnan2019dynamics} and {\textit{superdiffusion}}~\cite{PhysRevLett.106.220601, bertini2020finite,bulch2019superdiffusive}, has also been observed. The mechanisms underlying anomalous diffusion occurs remain an active open question.

The phenomenon of spin superdiffusion in the quantum Heisenberg spin chain, discovered in \cite{PhysRevLett.106.220601},
has recently been confirmed in a number of numerical studies  with tensor network simulations~\cite{lzp, PhysRevLett.122.210602, dupont_moore},
%Spin superdiffusion and Kardar-Parisi-Zhang scaling~\cite{kpz, quastel,Takeuchi2011} in the quantum Heisenberg spin chain were %discovered in tensor network simulations~\cite{PhysRevLett.106.220601, lzp, PhysRevLett.122.210602, dupont_moore}
and then addressed~\cite{idmp, gv_superdiffusion, PhysRevLett.123.186601, gvw, vir2019, dmki} using the framework of generalized hydrodynamics (GHD)~\cite{Doyon, Fagotti, SciPostPhys.2.2.014, PhysRevLett.119.020602,  BBH0, BBH,PhysRevLett.119.020602, GHDII, doyon2017dynamics, solitongases,PhysRevLett.119.195301,2016arXiv160408434Z, PhysRevB.96.081118,PhysRevB.97.081111, PhysRevLett.120.164101, dbd1, ghkv, dbd2, gv_superdiffusion, agrawal2019,Balasz,gvw, horvath2019euler, PhysRevB.100.035108,2019arXiv190601654B,10.21468/SciPostPhys.8.3.041,ruggiero2019quantum,friedman2019diffusive,bastianello2020generalised}, which extends hydrodynamics to integrable systems. The observed anomalous diffusion was initially attributed to particular properties of interacting quasiparticle excitations in the Heisenberg chain~\cite{idmp, gv_superdiffusion, PhysRevLett.123.186601}. More recent studies, however, uncovered the presence of universal KPZ dynamics in a wide class of quantum~\cite{dupont_moore} and classical~\cite{PhysRevE.100.042116,1909.03799,2003.05957} Hamiltonian systems, together with other types of superdiffusion \cite{PhysRevB.99.140301,Misguich2019}.
These include, among others, models which are directly relevant for cold atoms experiments such as the Fermi-Hubbard chain \cite{Schneider:2012aa,aidelsburger2018}. {At the same time, even if the KPZ equation was originally introduced to describe classical stochastic growth phenomena \cite{kpz}, its large dynamical universality class has recently incorporated also noisy quantum systems, as random unitary models \cite{Nahum2017} and spin chains with noise \cite{1912.08458,2001.04278}.}

Stimulated by previous observations, Refs.~\cite{vir2019, dmki} suggested that absence of normal diffusion originates from the long-wavelength fluctuations of local conserved charges associated with the non-Abelian continuous symmetry of the model.
A common theme that emerges from all of these studies is that the excitations responsible for the observed anomalous spin diffusion in the Heisenberg chain are interacting long-wavelength spin fluctuations: either a thermal gas of ``giant quasiparticles''~\cite{gv_superdiffusion, PhysRevLett.123.186601} described by GHD equations or, alternatively, ``soft gauge modes''
that conventional GHD cannot capture~\cite{vir2019, dmki}.
These pictures have complementary advantages: the GHD approach is microscopic, but has not so far been able to reproduce the emergence of the KPZ scaling function, whereas the latter is field-theoretical phenomenological approach which offers a plausible derivation of the KPZ equation.

\begin{figure}[!t]
\begin{center}
\includegraphics[width = 0.47\textwidth]{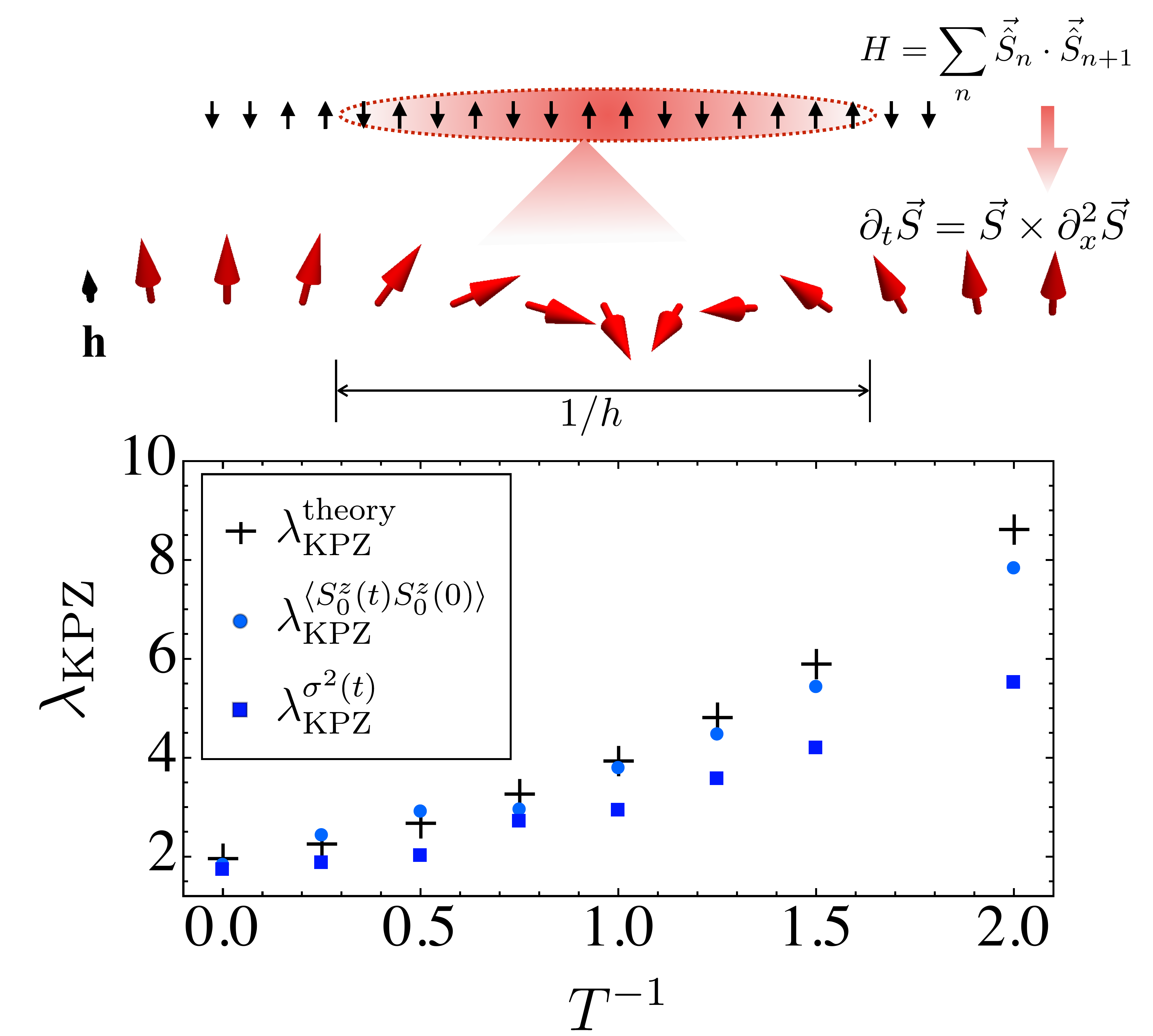}
\caption{Upper panel:  given an external small magnetic field $h \ll 1$, the quantum-classical mapping identifies quantum eigenstates with large quantum numbers $s \propto 1/h$ of magnons with soliton waves in a classical Landau-Lifshitz magnet. %The finite-density gas of such solitons has a diverging conductivity and dynamical response in the KPZ universality class. }
Lower panel: temperature ($T$)-dependent coefficient of the KPZ nonlinearity predicted by the self-consistent theoretical approach, compared with numerical results from tDMRG (see also caption of Fig. \ref{Fig:plot1}). At finite times, the spin profile is not precisely of the KPZ functional form as in eq. \eqref{eqKPZ}; thus the value of the inferred coefficient largely depends on which quantity one uses to extract to the KPZ prediction, either the spin auto-correlation or its variance.}
\label{fig1}
\end{center}
\end{figure}

In the present work, we elucidate the microscopic nature of spin superdiffusion by identifying the ``giant quasiparticles'' of the 
Heisenberg model with classical soliton solutions of the Landau-Lifshitz equation~\cite{lrt}.
We achieve this systematically through an explicit semi-classical scaling limit of the thermodynamic Bethe ansatz equations, thereby providing the missing link between the GHD approach and the proposal of Ref.~\cite{vir2019}.
This allows us to predict not only the correct exponent for superdiffusion but also the numerical value of the temperature-dependent coupling constant of the emergent KPZ dynamics (Fig.~\ref{fig1}). We thus show how for integrable classical or quantum isotropic ferromagnets, the data entering the coarse-grained KPZ equation can be derived microscopically.

\textbf{\textit{Model}}. 
We consider the spin-$\frac{1}{2}$ Heisenberg Hamiltonian for a chain of size $L+1$
\begin{equation} \label{eqH}
\hat{H} = J \sum_{n=-L/2}^{L/2} \vec{\hat{S}}_n \cdot \vec{\hat{S}}_{n+1},
\end{equation}
with $\vec{\hat{S}}_n=(S^x_n,S^y_n,S^z_n)$, and $\hat{S}^\alpha_n$ denoting spin-$\frac{1}{2}$ operators at site $n$.
In what follows, we set $J=1$. %We are interested in finite-temperature  spin transport in this model, and
We focus on the spin \textit{dynamical structure factor} $\langle \hat{S}^z_n(t) \hat{S}^z_0(0) \rangle$ %(connected correlation function) 
in the thermodynamic limit $L \to \infty$, at thermal equilibrium with \textit{finite temperature} $T>0$ and \emph{zero} magnetic field.
There is now ample numerical evidence that the structure factor at {late} times $t \gg 1$ follows the KPZ scaling form % \Jacopo{, instead of displaying standard \Romain{scaling}, is described by the KPZ dynamical universality class at late times}
\begin{equation} \label{eqKPZ}
 \langle \hat{S}^z_n(t) \hat{S}^z_0(0)\rangle \simeq \frac{\chi}{(\lambda_{\rm KPZ}\: t )^{2/3}} f_{\rm KPZ}\left( \frac{n}{(\lambda_{\rm KPZ}  \: t)^{2/3}}\right),
\end{equation}
where $\chi = \sum_{n=-L/2}^{L/2}  \langle \hat{S}^z_n(0) \hat{S}^z_0(0)\rangle$ is the static spin susceptibility, $f_{\rm KPZ}$ is the KPZ scaling function~\cite{Prahofer2004,quastel}, and $\lambda_{\rm KPZ}$ is the \emph{KPZ constant}: a temperature and model-dependent coupling parameter of the emergent KPZ equation describing the hydrodynamics of the spin field.
The exponent $2/3$ can be extracted from GHD via a self-consistent argument~\cite{gv_superdiffusion, PhysRevLett.123.186601, gvw}, but the scaling function and $\lambda_{\rm KPZ}$ cannot. Moreover, the framework of nonlinear fluctuating hydrodynamics~\cite{Das:2019aa,Spohn2014}, {which has been used to derive KPZ equations in other contexts}, \textit{does not} apply straightforwardly in this situation.

Recent numerical results  \cite{PhysRevE.100.042116}, supplemented by theoretical arguments of refs.~\cite{vir2019,dmki}, have
presented evidence that the same universal KPZ scaling also occurs at finite temperatures in classical integrable spin chains invariant under $SO(3)$ rotations, whose continuum long-wavelength theory is governed by the Landau-Lifshitz (LL) equation
\begin{equation}\label{eq:LLeq}
\partial_t \vec{S} = J_{\rm cl} \ \vec{S}\times \partial_x^2 \vec{S},
\end{equation}
where $\vec{S}\equiv \vec{S}(x,t)$ is a classical spin field of unit length $|\vec{S}|=1$ on the continuum line $x \in \mathbb{R}$.
In this light, it is reasonable to expect that such emergent behavior is a manifestation of a quantum-classical correspondence where certain degrees of freedom in the quantum chain are intrinsically classical in nature and behave according to \eqref{eq:LLeq}, as proposed, e.g., in Ref.~\cite{vir2019}.

Here, we isolate the excitations relevant for KPZ dynamics. Since these turn out to be bound states of elementary magnonic excitations whose size and quantum numbers diverge as the local magnetization vanishes, we dub them ``giant quasiparticles''.
Our picture, combined with simple kinetic arguments, yields quantitative predictions for the $\lambda_{\mathrm{KPZ}}$,
and elucidates how a finite thermal density of giant quasiparticles in the spectrum of the quantum chain leads to a \textit{thermal gas} of classical solitons of the Landau-Lifshitz field theory \eqref{eq:LLeq}.

\textbf{\textit{Computing the KPZ constant}}. The KPZ coupling constant of the quantum Heisenberg model can be computed from the
following procedure. First, we consider a thermal Gibbs state with the addition of a small magnetic field $2h$, which introduces the additional term  $ -2hT \sum_i \hat{S}_i^z$ to $\hat{H}$.
Given that the model~\eqref{eqH} is integrable, spin dynamics splits into two channels; a ballistic piece with spectral (Drude) weight vanishing at zero field, and a diffusive part with spin diffusion constant \emph{diverging} as $D(h) = D_0/h$ in the $h \to 0$ limit~\cite{PhysRevLett.123.186601, gvw}. Both transport coefficients admit closed-form expressions as sums over quasiparticles, labelled by a discrete label $s \geq 1$ (pertaining to the quantized magnetization of magnon excitations) and a continuous rapidity label $\theta \in (-\infty, \infty)$ which parametrizes their quasimomenta $p_{s}(\theta)$. The spin diffusion constant assumes a spectral decomposition~\cite{dbd2,PhysRevLett.123.186601,medenjak2019diffusion,doyon2019diffusion}
\begin{equation}\label{eqD}
D = \sum_{s \geq 1} \int_{-\infty}^\infty d\theta \, D_s(\theta),
\end{equation}
which we will use below to determine $D_0$.

The second step of our procedure consists of regularizing the divergence of $D(h)$ by accounting that the net magnetization observed by a quasiparticle that has traveled a distance $\ell$ is not precisely zero, but instead has a residual value $h(\ell)$ set by thermal magnetization fluctuations over the scale $\ell$. As noted previously in Ref.~\cite{gv_superdiffusion}, the motion of the giant quasiparticles that dominate spin transport is primarily diffusive, so $\ell$ is itself self-consistently set by $h(\ell)$.
These equations relating $\ell$ and $h(\ell)$ now permit for a quantitative analysis of superdiffusion in terms of $D_0 = \lim_{h \to 0^+} h D$. For $h \ll 1$, this can be though of as the effective field originating from thermal fluctuations, namely $
h^2 = {m^2}/{(4 \chi)^2}   
$, with $m^2$ the local spin susceptibility in a interval of size $\ell$, $m^2= {4 \chi}/{\ell}$. We then infer $h= 1/\sqrt{4 \chi \ell}$,
and the length-scale $\ell$ can be fixed self-consistently at small, finite $h$ by $\ell^2 = 2 D\, t = 2 D_0  \sqrt{4 \chi \ell}\, t$. This yields $\ell =  (2 D_0  \sqrt{4 \chi}  t)^{2/3}$, and  combining gives finally
\begin{equation}
D(t)  =  2^{5/3}  D_0^{4/3} \chi^{2/3} t^{1/3} + \dots
\end{equation}
This simple argument already suffices to predict anomalous diffusion with dynamical exponent $z=3/2$. Remarkably, it also predicts the value of the prefactor. Even though such an approach is arguably heuristic, we wish to emphasize that a similar argument correctly predicts the exact form~\cite{gv_superdiffusion} of the diffusion constant~\eqref{eqD}
for the easy-axis XXZ spin chain, which has been computed by other means~\cite{dbd2,PhysRevLett.123.186601,medenjak2019diffusion,doyon2019diffusion},
so it should be taken seriously.
To extract $\lambda_{\rm KPZ}$ defined in eq.~\eqref{eqKPZ}, we compare the variance of the spin profile $\sigma^2$ to the variance computed from the KPZ prediction~\eqref{eqKPZ}. At any \textit{finite} $t$ there is a finite (diverging) diffusion constant $D(t) \sim t^{1/3}$,  which we can define in terms of the spin variance as $\sigma^2= \chi 2 D(t) t$. This readily implies that the full
temperature-dependent KPZ constant $\lambda_{\rm KPZ} \equiv \lambda_{\rm KPZ}(T)$ is given by
\begin{equation} \label{eqLambda}
\lambda_{\rm KPZ}(T) = 4 D_0(T) \sqrt{\chi(T)}/\sigma_{\rm KPZ}^{3/2}.
\end{equation}
Here $\sigma_{\rm KPZ}^2$ is the variance of the KPZ function $\sigma_{\rm KPZ}^2=\int du \ u^2 f_{\rm KPZ}(u) \approx 0.510523$. Let us stress again that the above argument does not predict the KPZ scaling function, but it does fix $\lambda_{\rm KPZ}$ as a function of temperature.

\begin{figure}[!t]
\includegraphics[width=0.47\textwidth]{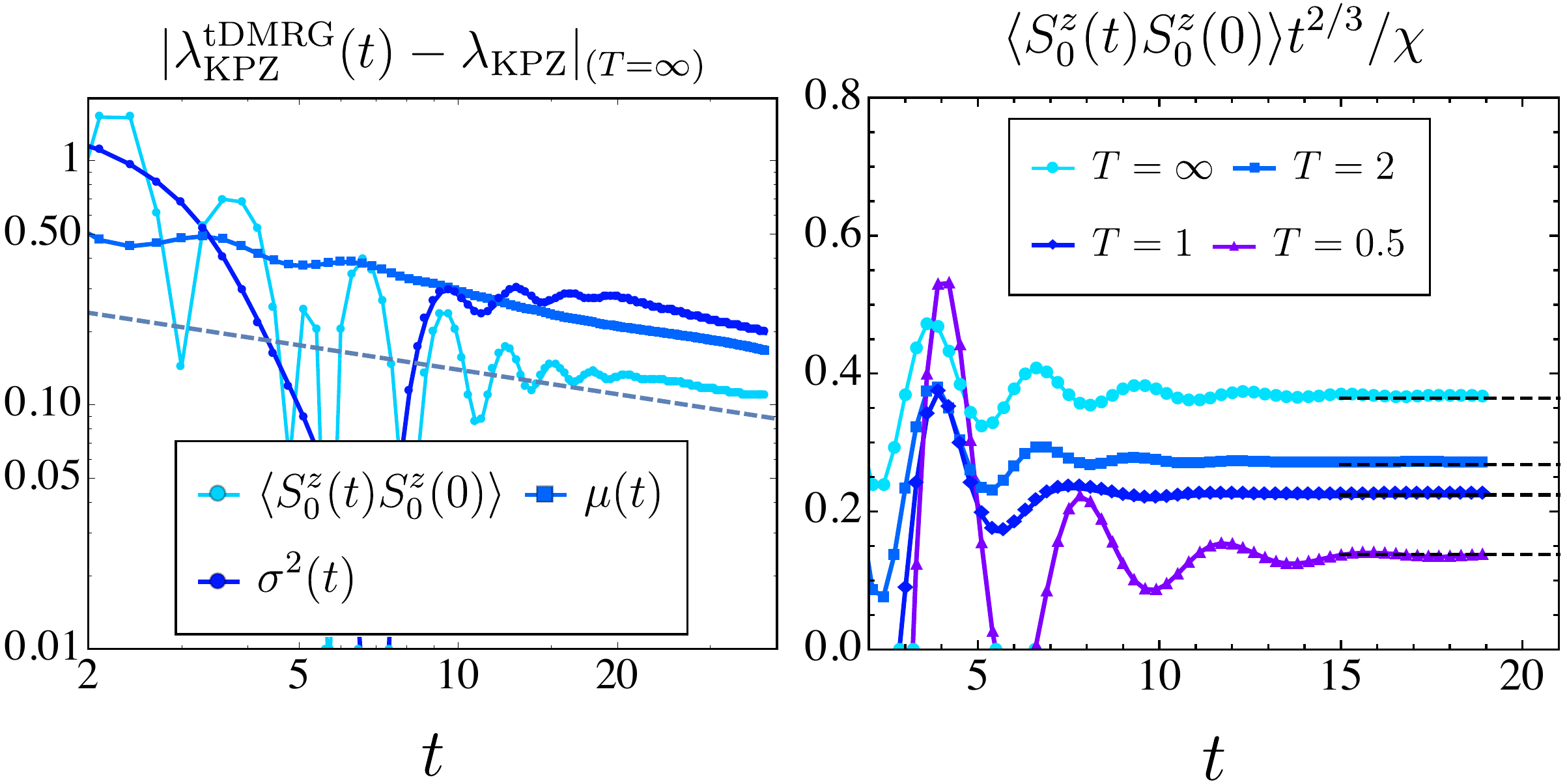}
\caption{ \textit{Left}: Log-Log plot of the $\lambda_{\rm KPZ}$ computed from tDMRG numerical simulations in a Heisenberg chain at infinite temperature $T=\infty$, minus our theoretically predicted value $\lambda_{\rm KPZ}(T = \infty) = 1.9265 \ldots$, as function of the numerical simulation times $t$ (in unit of spin coupling $J$). Dashed gray line represents $t^{-1/3}$. \textit{Right}:  Spin auto-correlation as function of time multiplied by $t^{2/3}$. Convergence to KPZ scaling is reached at $t \sim O(10)$ for all considered temperatures $T$.   }
\label{Fig:plot1}
\end{figure}

\textbf{\textit{Giant quasiparticles as classical soft solitons}}. 
Our central result is the explicit form \eqref{eqLambda} for $\lambda_{\rm KPZ}$ in terms of parameter $D_0(T)$. Now we explain
how to explicitly compute it.
The following calculation also demystifies the nature of the ``giant quasiparticles'' responsible for superdiffusion: following previous work~\cite{gv_superdiffusion, PhysRevLett.123.186601, gvw} we anticipate that these are semiclassical quasiparticles carrying large amount of spin $s \sim 1/h$, i.e., macroscopically large bound states of magnons which belong to the low-energy spectrum of the Heisenberg chain.
Such states, first described in Refs.~\cite{Sutherland1995,PhysRevLett.85.2813}, have received a great deal of attention in the study 
of gauge-string dualities \cite{Minahan2003,Arutyunov2007,Roiban2007,Minahan2008,beisert2012review}.
As explicitly shown in~\cite{Kazakov2004,Bargheer2008}, semiclassical eigenstates manifest themselves (at the classical level) as solutions to the continuum Landau-Lifshitz model~\cite{Takhtajan1977}.

Our objective here is however not to describe individual classical spin-field configurations but rather find a classical interpretation for the giant quasiparticles immersed in a thermal background.
To this end, we identify an appropriate semi-classical limit directly at the level of the thermodynamic Bethe ansatz (TBA) equations.
We shall see that this will lead us directly to the classical counterpart of the GHD equations where,
remarkably, the small magnetic field $h$ will play the role of an effective \textit{Planck constant}. With this in mind, we introduce a rescaled rapidity $u = \theta h$ and rescaled quasiparticle magnetization $\xi= s\,h$.
In the limit $h \to 0^+$, we can convert the sum over $s$ in eq.~\eqref{eqD} into an integral, in this way obtaining a fully	
classical expression for $D_0 \equiv  D_0^{\rm cl}$, with
\begin{equation}
D_0^{\rm cl }= \int_0^{+\infty}\!\! d\xi \int_{-\infty}^{+\infty}\!\! du D(\xi,u),
\end{equation}
and  where $ D(\xi,u) =  \lim_{h \to 0^+} \frac{1}{h} D_{\xi/h}(u/h)$ is a finite quantity: $D_0$ is thus fully determined by quasiparticles with $s \to \infty$ in the limit $h \to 0^+$, with $\xi=s\,h$ kept fixed.

Next, we consider the scattering phase shifts between two quasiparticles with spin indices $s$ and $s'$ with relative rapidity $\theta$. Here we quote the result of \cite{Takahashi1971,Takahashi}, $T_{s,s'}(\theta) = (1 - \delta_{s s'}) a_{|s-s'|}(\theta) + 2  a_{|s-s'|+2}(\theta) +\dots +  2a_{s+s'-2}(\theta)+ a_{s+s'}(\theta)$, with $a_s(\theta) \equiv \frac{1}{2\pi}\partial_{\theta}p_{s}(\theta) = \frac{1}{2\pi}  \frac{4s}{  s^2 + 4 \theta^2 }$. Upon rescaling of parameters $\theta \to u / h$ and $s\to \xi/h$, the net
phase shift of all the constituent magnons can be resummed into an integral
$T_{s,s'}=\int_{h|s-s'| }^{ h(s+s') } \frac{d \zeta}{2\pi} \frac{4\zeta}{\zeta^2 + 4 u^2 }$, up to ${\cal O}(h)$ corrections.
This readily provides an effective scattering kernel for the giant quasiparticles
$T^{\rm giant}_{\xi,\xi'}(u) = \lim_{h \to 0^+} T_{\xi/h,\xi'/h} \left(u/h\right)$, reading explicitly
\begin{equation}\label{eq:scatteringshift}
T^{\rm giant}_{\xi,\xi'}(u) = \frac{1}{\pi} \log \frac{4 u^2 + (\xi+\xi')^2}{4 u^2 + (\xi-\xi')^2} .
\end{equation}
In this expression one can recognize the scattering kernel -- the differential scattering phase of the two-body $\mathcal{S}$-matrix -- ascribed to an elastic collision of two Landau-Lifshitz solitons~\cite{Takhtajan1977,ludwigfaddeev2007} characterized by pairs of 
action variables $(u_{1},\xi)$ and $(u_{2},\xi')$ with $u=u_{1}-u_{2}$.
All the remaining thermodynamic state functions pertaining to the giant quasiparticles can be obtained in a similar manner
by rescaling the analogous quantities in the Heisenberg chain.
We will need the following standard TBA concepts: an equilibrium state is uniquely characterized by a density $\rho_s(\theta)$ of
quasiparticles with quantum numbers $(s,\theta)$, the available density of states $\rho^{\mathrm{tot}}_s(\theta)$ and
the associated Fermi filling fractions $n_s(\theta) \equiv \rho_s(\theta)/\rho^{\mathrm{tot}}_s(\theta)$. Finally, interactions ``dress'' the group velocity and magnetization (along with other local charges) carried by quasiparticles; we denote these $v^{\mathrm{eff}}_s(\theta)$ and $m_s^{\mathrm{dr}}(\theta)$ respectively.

Writing the Fermi filling functions of the quasiparticles as $n_s(\theta) = (1 + \eta_s(\theta))^{-1}$, we have  $\eta_{\xi/h}(u/h) \to  \eta(\xi,u)/h^2$, implying vanishing occupations $n_s \sim h^2 \eta^{-1}(\xi,u)$ and \textit{emergent classical statistics} for these modes.
The rescaled ratio $\eta(\xi,u)$ is interpreted as a Boltzmann weight which obeys a two-dimensional Fredholm-type integral equation~\cite{suppmat}
\begin{align}\label{eq:etaclassical}
&\log {\eta} (\xi,u)  =  2 \log h +  2 \xi -  \frac{h}{T} e^{\rm giant}(\xi,u) \nonumber \\& \qquad + \int_0^{+\infty} [d \xi']_h \int_{-\infty}^{+\infty} dv \, T^{\rm giant}_{\xi,\xi'}(u-v) [{{\eta} (\xi',v)}]^{-1},
\end{align} 
where we have introduced a regularized integral
$\int_0^{+\infty} [d\xi]_h g(\zeta) \equiv \int_0^{+\infty} d \zeta g(\xi) - \frac{h}{2} \lim_{\xi \to 0^+} g(\xi)$ for any function $g(\xi)$ and bare energy $e^{\rm giant}(\xi,u)=2\xi/(\xi^2 + 4 u^2)$.
Equation \eqref{eq:etaclassical} can be interpreted as a \emph{semiclassical TBA equation} for a finite-density soliton gas.
Analogous integral equations, albeit without a regulator, have previously appeared in the context of classical thermodynamic soliton gases~\cite{Mertens1981,Bolterauer1981,Timonen1986,Sasaki1986,Bullough1990,Theodorakopoulos1991,Theodorakopoulos1995,2016arXiv160308628D,BDWY2018}.
We note however that eq.~\eqref{eq:etaclassical} only governs a particular scaling regime of classical \textit{``soft solitons''} with
low energy and large width.

Before we proceed with solving eq.~\eqref{eq:etaclassical}, we owe to clarify an important subtlety. Even though we are eventually only interested in the solution at $h=0$, the limit $h \to 0^+$ can be taken only after solving \eqref{eq:etaclassical}, as $h$ acts as a cut-off in the integral over the solitons' charge $\xi$. Similar integral equations can be also written for the densities of	
quasiparticles $\rho^{\rm tot}_{\xi/h}(u/h) \to h^2 \rho^{\rm tot}(\xi,u)$, the dressed rapidity-derivative of energy of the quasiparticle excitations $\varepsilon'_{\xi/h}(u/h) = h^3 \varepsilon'(\xi,u)$, and the dressed magnetization $m^{\rm dr}_{\xi/h}(u/h) = h^{-1} m^{\rm dr}(\xi,u)$~\cite{suppmat}. The dressed magnetization diverges as $h^{-1}$ in the  $h\to 0^+$ limit, while the velocity $v^{\rm eff}_{\xi/h}(u/h)=\varepsilon'_{\xi/h}(u/h)/(2 \pi \rho^{\rm tot}_{\xi/h}(u/h))$ vanishes as $\sim h$. It is also straightforward to check that these expressions are consistent with $D(\xi,u) =  \lim_{h \to 0^+} \frac{1}{h} D_{\xi/h}(u/h)$ converging to a finite function.
 
In the limit of infinite temperature,  $T=\infty$, dependence on parameter $u$ drops out of equation \eqref{eq:etaclassical},
which enables us to solve it exactly~\cite{Sasaki1986}. We find $\eta(\xi,u) = \sinh^{2}(\xi + h)$, in agreement with rescaling the exact analytical solution of the TBA equations at infinite temperature for the quantum chain~\cite{Takahashi}. Using this result, all other thermodynamic functions can also be obtained in a closed form, yielding
$D_0 (T \to \infty)= 5 \pi/27$~\cite{suppmat}. From equation \eqref{eqLambda} we thus deduce that
%
%\begin{equation}\label{eq:lambdaexact}
$\lambda_{\rm KPZ}(T = \infty) = 10 \pi/(27 \sigma_{\rm KPZ}^{3/2}) \approx 1.9265\dots$
%\end{equation}

\textbf{\textit{Numerical results}}. Solving the semi-classical TBA equation~\eqref{eq:etaclassical} at finite temperature $T$ is numerically challenging; in practice it is more convenient to solve the original quantum TBA equations and afterwards take the limit $D_0 = \lim_{h \to 0^+}h D(h) $ numerically. We compared our predictions to tDMRG calculations, see Fig.~\ref{fig1} and \cite{suppmat} for additional numerical data, by evolving a finite-temperature state \cite{Karrasch2012}, with fixed maximal bond dimension equal to $800$ and system size $L=140$ and computing the dynamical structure factor (DSF) $C_n(t)=\langle \hat{S}_{n}^z(t) \hat{S}_{0}^z(0) \rangle_T$ at finite temperature $T$.
Despite entanglement entropy growing linearly in time, we carry out computation up to times $t \sim 50 J$
and estimate the maximal error by comparing values of different observables.  In particular we extract the value of $\lambda_{\rm KPZ}$ at finite time by considering (given eq. \eqref{eqKPZ}): the auto-correlation, via $ \lambda_{\rm KPZ}^{\rm tDMRG}(t) = (t^{2/3}C_0(t)/(\chi f_{\rm KPZ}(0)))^{-3/2}$, the variance $\sigma^2(t) =   \sum_{n=-L/2}^{L/2} n^2 C_n(t)$, via $\lambda_{\rm KPZ}^{\rm tDMRG}(t) =(  t^{-4/3} \sigma^2(t)/(\chi \sigma_{\rm KPZ}^2))^{3/4}$ and the mean of the absolute value  $\mu(t) = \sum_{n=-L/2}^{L/2} |n| C_n(t)$ via similar relation. In the limit $t \to \infty$, all these values are expected to be equal and identify to $\lambda_{\rm KPZ}$. At the finite times accessible by the numerical simulation, we find an expected slow convergence towards the theoretically predicted value of $\lambda_{\rm KPZ}$, with corrections of order $t^{-1/3}$, 
consistently with other dynamical systems in the KPZ universality class \cite{Takeuchi2011,Ferrari2011} (Fig.~\ref{Fig:plot1}).

We find good agreement with our prediction \eqref{eqLambda}, especially at high temperature (Fig.~\ref{fig1}). At lower temperatures, however, various numerical estimators for $\lambda_{\rm KPZ}$ show some discrepancy, indicating that on the accessible time-scale the dynamical correlations have not yet relaxed sufficiently close to the asymptotic KPZ scaling form~\eqref{eqKPZ}.  We moreover observe that $\lambda_{\rm KPZ} \to \infty$ with decreasing temperature, suggesting that the classical KPZ dynamics only becomes valid on increasingly large spatio-temporal scales, whereas on shorter scales one can expect Luttinger liquid ballistic dynamics \cite{Karrasch2015,PhysRevB.66.144416} and spinon physics \cite{Mourigal2013}.

\textbf{\textit{Conclusion}}. We have traced  
the microscopic origin of anomalous spin transport in the quantum Heisenberg spin-$1/2$ chain
to the presence of giant quasiparticle eigenstates in its spectrum. These states admit a purely classical
interpretation as a thermal gas of soft classical solitons of the isotropic Landau-Lifshitz equation.
We established an explicit correspondence through the semi-classical scaling limit of the thermodynamic Bethe ansatz equations.
The Fermi factors of such giant quasiparticles are vanishingly small so they become effectively classical.

Our analysis unifies the complementary pictures of KPZ superdiffusion: the generalized hydrodynamics approach of Refs.~\cite{gv_superdiffusion, PhysRevLett.123.186601}, and the effective theory of Ref.~\cite{vir2019} which seemingly evades the conventional GHD description. In the language of GHD, one divides a system up into hydrodynamic cells of some fixed size, and constructs a thermal state within each cell. To construct such a state, one must specify both a ``pseudovacuum'' (i.e., a unit vector on the sphere which sets the direction of the net magnetization) and a quasiparticle distribution above this pseudovacuum. Ref.~\cite{vir2019} postulated a Landau-Lifshitz dynamics for long-wavelength spatial fluctuations of this pseudovacuum, arguing it cannot be captured by GHD modes. However, in light of our analysis, the distinction between such ``pseudovacuum fluctuations'' and quasiparticles is only superficial as is depends on the cut-off: pseudovacuum fluctuations are nothing but giant quasiparticles that extend beyond the scale of a hydrodynamic cell, and can indeed be described within GHD. With that, we confirm the previous suggestion~\cite{vir2019} that superdiffusion in the Heisenberg spin chain is due to low-energy degrees of freedom that obey an emergent Landau-Lifshitz equation; the quantum and classical systems share the same hydrodynamic description in terms of a stochastic Burgers (or equivalently KPZ) equation. Our explicit derivation provides the microscopic input for the KPZ equation, permitting to determine the temperature dependence of its coupling
constant (in good agreement with numerical results); moreover, it establishes the universal nature of the low-energy solitons that cause superdiffusion. We expect the explicit mapping to a classical model to enable efficient numerical simulations that should quantitatively address important questions such as the fate of superdiffusion away from strict integrability, see~\cite{dmki}.

Our results can be straightforwardly generalized to other integrable spin/charge models where KPZ scaling is also expected,
including the spin-$S$ integrable chains, integrable models of higher-rank symmetry~\cite{2003.05957}
and Fermi-Hubbard chains~\cite{idmp}.
A separate interesting direction for future work would be to understand the crossover from Luttinger liquid physics to KPZ dynamics at low temperature.

\begin{acknowledgments}

{\textit{\textbf{Acknowledgments}}}. We are very grateful and indebted to Benjamin Doyon, Takato Yoshimura, Tomohiro Sasamoto for inspiring discussions on the  semiclassical TBA equations and collaboration on the KPZ problem in the XXX chain; to Marko Medenjak and Brayden Ware for collaborations on closely related topics; to Utkarsh Agrawal for early collaboration on the numerical solutions to the TBA equations of the Heisenberg chain;  and to Vir Bulchandani for numerous stimulating discussions.   We thank the International Centre for Theoretical Sciences (ICTS) and the program  ``Thermalization, Many body localization and Hydrodynamics'' (Code: ICTS/hydrodynamics2019/11) where this project was initiated. The MPS-based tDMRG simulations were performed using the ITensor Library~\cite{ITensor}. This work was supported by the National Science Foundation under NSF Grant No. DMR-1653271 (S.G.),  the US Department of Energy, Office of Science, Basic Energy Sciences, under Early Career Award No. DE-SC0019168 (R.V.),   the Alfred P. Sloan Foundation through a Sloan Research Fellowship (R.V.), the Research Foundation Flanders (FWO, J.D.N.), and the Slovenian Research Agency
(ARRS) program P1-0402 (E.I.).

\end{acknowledgments}

\bibliography{refs}

%%%%%%%%%%%%%%%%%%%%%% Include the appendix %%%%%%%%%%%%%%%%%%%%%%%%%%
%

% Page by page for Arxiv

\bigskip

\onecolumngrid
\newpage



\section*{Supplementary material}

# Supplemental Material for
# "Superdiffusion from emergent classical solitons in quantum spin chains"

Jacopo De Nardis,[1] Sarang Gopalakrishnan,[2] Enej Ilievski,[3] and Romain Vasseur[4]

[1]*Department of Physics and Astronomy, University of Ghent, Krijgslaan 281, 9000 Gent, Belgium*
[2]*Department of Physics and Astronomy, CUNY College of Staten Island,
Staten Island, NY 10314; Physics Program and Initiative for the Theoretical Sciences,
The Graduate Center, CUNY, New York, NY 10016, USA*
[3]*Faculty for Mathematics and Physics, University of Ljubljana, Jadranska ulica 19, 1000 Ljubljana, Slovenia*
[4]*Department of Physics, University of Massachusetts, Amherst, Massachusetts 01003, USA*

(Dated: May 20, 2020)

## I. SPIN DIFFUSION CONSTANT

### A. Kinetic theory derivation of the spin diffusion constant

We first recall how one can compute the spin diffusion constant in the gapped regime $\Delta \geq 1$ of the anisotropic XXZ spin chain

$$H = \sum_{n=-L/2}^{L/2} (\hat{S}_n^x \hat{S}_{n+1}^x + \hat{S}_n^y \hat{S}_{n+1}^y + \Delta \hat{S}_n^z \hat{S}_{n+1}^z), \tag{1}$$

using a kinetic picture based on thermal fluctuations of the dressed magnetization. We follow the approach of Ref.[1], generalized to finite temperature. In the presence of a small magnetic field $h$, the variance of the spin-structure factor is scaling ballistically with the Drude weight[2,3]

$$\sigma^2 \equiv \sum_{n=-L/2}^{L/2} n^2 \langle S_n^z(t) S_0^z(0)\rangle = t^2 \sum_s \int d\theta \rho_s(\theta)(1-n_s(\theta))(v_s^{\rm eff}(\theta)m_s^{\rm dr}(\theta))^2. \tag{2}$$

As $h \to 0$ (or half-filling, in the equivalent fermionic language), the dressed magnetization vanishes, giving rise to normal spin diffusion. To see this, we write down the effective dressed magnetization felt by a $(s,\theta)$-quasiparticle propagating with an effective velocity $v_s^{\rm eff}(\theta)$,

$$(m_s^{\rm dr}(\theta))^2 = \frac{1}{2}\left.\partial_h^2(m_s^{\rm dr})^2\right|_{h=0} h^2 + \ldots, \tag{3}$$

where $h^2$ can be though of as the effective field originating from thermal fluctuations:

$$h^2 = \frac{1}{(4\chi)^2}m^2 = \frac{1}{4\chi|v_s^{\rm eff}(\theta)t|}. \tag{4}$$

Here we have taken the fluctuations

$$m^2 = \left\langle \left(\frac{1}{\ell}\sum_{n=n_0}^{n_0+\ell} \sigma_n^z\right)^2\right\rangle = \frac{4\chi}{\ell}, \tag{5}$$

over a distance $\ell = |v_s^{\rm eff}(\theta)t|$, and the factor of 4 in front of $\chi$ is to match the usual Bethe Ansatz notation. This yields

$$\sigma^2 = t\sum_{s\geq 1}\int d\theta \rho_s(\theta)(1-n_s(\theta))\left|v_s^{\rm eff}(\theta)\right|\frac{1}{8\chi}\left.\partial_h^2(m_s^{\rm dr})^2\right|_{h=0}. \tag{6}$$

To identity the diffusion constant, we note that the variance of the spin structure factor should scale as $\sigma^2 = \chi 2Dt$. We thus get

$$D = \sum_{s\geq 1}\int d\theta \rho_s(\theta)(1-n_s(\theta))\left|v_s^{\rm eff}(\theta)\right|\frac{1}{16\chi^2}\left.\partial_h^2(m_s^{\rm dr})^2\right|_{h=0}. \tag{7}$$

Introducing the filling factor $\nu = 4m = 4\chi h$, this can in turn be rewritten as

$$D = \sum_{s\geq 1} \int d\theta \rho_s(\theta)(1 - n_s(\theta)) \left|v_s^{\text{eff}}(\theta)\right| \partial_\nu^2 (m_s^{\text{dr}})^2\big|_{\nu=0}. \tag{8}$$

This simple argument predicts a diffusion constant in agreement with other approaches[4,5]. The argument used in the main text to derive the KPZ coupling $\lambda_{\text{KPZ}}$ is similar, and relies on the same physical assumptions. This expression for the diffusion constant is valid for $\Delta \geq 1$, and diverges in the isotropic limit $\Delta \to 1$.

### B. "Magic formula" and curvature of the Drude self-weight

The formula (8) derived above agrees with the general expression derived from thermal form factors in Refs.[4,6], combined with the "magic formula" of Ref.[5]. This simplified expression is strictly valid in the XXZ spin chain with $\Delta > 1$ and at the isotropic point $\Delta = 1$, in the absence of magnetic field ($h = 0$). At the isotropic point, this predicts a divergence $D = D_0/h$ with a coefficient $D_0$ that evaluate in the main text (see also below), with $D_0 = \frac{5\pi}{27}$ at infinite temperature. Here we note that this limit is rather subtle: for example, eq. (8) has been interpreted in Ref.[5,7] as the curvature of the "Drude self-weight"

$$\tilde{D} = \frac{1}{16\chi^2} \lim_{h\to 0} \frac{\partial^2}{\partial h^2} \sum_{s\geq 1} \int d\theta \rho_s(\theta)(1 - n_s(\theta)) \left|v_s^{\text{eff}}(\theta)\right| (m_s^{\text{dr}})^2. \tag{9}$$

This expression differs from eq. (8) only from the derivatives being taken outside of the sum over strings (and the integral over rapidity). While those two expressions agree in the case where the diffusion constant is finite, we find that $\tilde{D}$ is only a lower bond to $D$ at the isotropic point. In particular, it is easy to show that it diverges as

$$\tilde{D} = \tilde{D}_0/h, \tag{10}$$

as $h \to 0$ with

$$\tilde{D}_0 = \frac{4}{\pi} \int_0^\infty dx \frac{(-1 + x \coth x)^2 (1 + x \coth x)}{x^2 \sinh^2 x} \approx 0.465218. \tag{11}$$

at infinite temperature. We see that $\tilde{D}_0 < D_0 = 5\pi/27$ at infinite temperature.

Evidently extracting the precise nature of the divergence of the diffusion constant as $h \to 0$ is rather subtle. To check the validity of the expression (8), we go back to the full expression of the diffusion matrix derived in Refs.[4,6]. Using the notations of Ref.[4], the $DC$ matrix is given by

$$DC = R^{-1} \rho^{\text{tot}} \widetilde{D} n(1-n) R^{-1}, \tag{12}$$

where $C$ is the susceptibility matrix. Evaluating this expression for the spin $m_s = s$ gives

$$[DC]_{\text{spin}} = (m, DCm) = \sum_s \int d\theta \rho_s(\theta)(1 - n_s)(m_s^{\text{dr}})^2 w_s(\theta) + \ldots \tag{13}$$

where

$$w_s(\theta) = \lim_{s_{\max} \to \infty} \sum_{s'=1}^{s_{\max}} \int d\alpha \rho_{s'}(\alpha)(1 - n_{s'})|v_s(\theta) - v_{s'}(\alpha)|(T^{\text{dr}}(\theta,\alpha)_{s,s'}/\rho_s^{\text{tot}}(\theta))^2, \tag{14}$$

in terms of the dressed kernel $T^{\text{dr}}$. The dots in (13) correspond to off diagonal terms that can be argued to vanish as $h \to 0$[4]. For the XXZ spin chain ($\Delta > 1$) as $h \to 0$, the sum over $s$ is pushed to larger and larger $s$ due to the vanishing term $(m_s^{\text{dr}})^2$. Since $w_\infty$ is well defined at any $\theta$, we have

$$[DC]_{\text{spin}} = w_\infty \sum_s \rho_s(1 - n_s)(m_s^{\text{dr}})^2 = w_\infty \chi, \tag{15}$$

so the diffusion constant is given by $w_\infty$.

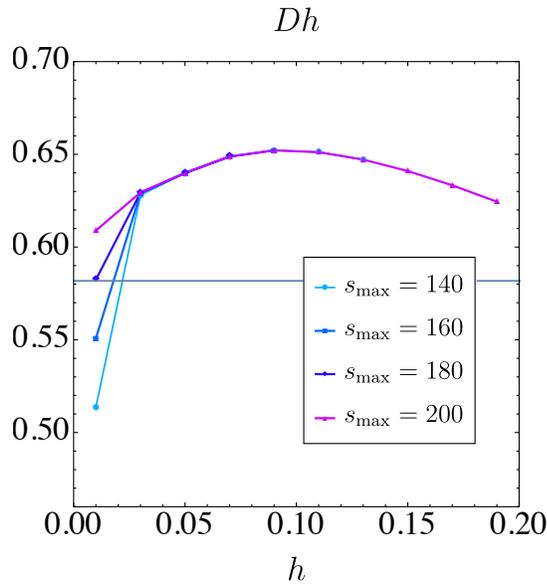

FIG. 1: Plot of the spin diffusion constant at infinite temperature, computed from equation (13) as the external field is lowered to zero and for different maximal string number $s_{\max}$. In the limit $s_{\max} \to \infty$ the value $\lim_{h \to 0} Dh = 5\pi/27$ (22) (blue line) is recovered.

However, we remark that this limit is not defined in the XXX model since there is no convergence for any $\theta \gg 1$. So the correct diffusion formula in this case is

$$[D]_{\text{spin}} = \chi^{-1} \lim_{s_{\max} \to \infty} \sum_{s=1}^{s_{\max}} \int d\theta \rho_s(\theta)(1-n_s)(m_s^{\text{dr}})^2 w_s(\theta) + \mathcal{O}(h), \qquad (16)$$

with

$$\chi = \sum_s \int d\theta \rho_s(\theta)(1-n_s)(m_s^{\text{dr}})^2, \qquad (17)$$

where the $\mathcal{O}(h)$ terms come from off-diagonal elements of the diffusion matrix, that should vanish in the zero field limit. We have evaluated this formula numerically for finite $h$ and finite $s_{\max}$ (see Fig. 2), and found a divergence of the diffusion constant compatible with $D = D_0/h$ and $D_0 = 5\pi/27$ at infinite temperature, as predicted from the simplified formula (8).

## II. SEMICLASSICAL THERMODYNAMICS BETHE ANSATZ EQUATIONS FOR THE GIANT QUASIPARTICLES

Consider now the isotropic Heisenberg chain, with the addition of external magnetic field $-2hT \sum_i \hat{S}_i^z$. In the limit $h \to 0$, the spin diffusion constant diverges as $D = D_0/h$ in a thermal Gibbs ensemble at finite temperature $T$. Our aim is to extract the thermodynamic quantity $D_0$. We begin with the spectral resolution

$$D = \sum_{s \geq 1} \int d\theta D_s(\theta), \qquad (18)$$

where

$$D_s(\theta) = \frac{1}{(4\chi)^2} \rho_s(\theta)(1-n_s(\theta))|v_s^{\text{eff}}(\theta)|\partial_h^2 (m_s^{\text{dr}}(\theta))^2. \qquad (19)$$

In terms of rescaled variables $\xi = sh$ and $u = s\theta$, we find at small $h$

$$D = \frac{2}{h} \sum_{\xi/h \geq 1} \int_0^\infty du D_{\xi/h}(u/h), \tag{20}$$

and we have the following well-defined limit

$$D_0 = 2\int_0^\infty d\xi \int_0^\infty du \lim_{h\to\infty} \frac{1}{h} D_{\xi/h}(u/h) = 2\int_0^\infty d\xi \int_0^\infty du D(\xi, u). \tag{21}$$

Using the known analytical expressions of the TBA functions $\rho_s(\theta)$, $n_s(\theta)$, $v_s^{\text{eff}}(\theta) = \varepsilon'_s(\theta)/(2\pi\rho_s^{\text{tot}}(\theta))$ and $m_s^{\text{dr}}(\theta)$ for the infinite temperature Gibbs state, for example written in Ref. 8 and reported below, we find

$$D_0(T \to \infty) = \frac{1}{2}\int_0^\infty d\xi \int_0^\infty du \frac{32u\xi^4}{(\sinh\xi)^2} \frac{\left(-4u^2 + \xi\left(4u^2 + \xi^2\right)\coth\xi + 3\xi^2\right)}{9\pi\left(4u^2 + \xi^2\right)^3} \frac{1}{(2\chi)^2}, \tag{22}$$

$$= \int_0^\infty d\xi \left(\frac{2\xi}{\sinh\xi}\right)^2 \frac{(\xi(\coth\xi)+1)}{9\pi} = \frac{5\pi}{27}. \tag{23}$$

This expression, valid at infinite temperature, is an exact prediction. Although we have obtained it here by rescaling the known solution of the TBA equations for the Heisenberg quantum chain, we show in the following that coefficient $D_0$ can be expressed in terms of rescaled thermodynamic quantities for arbitrary temperature, namely

$$D_0 = \frac{1}{(4\chi)^2}\int_0^\infty d\xi \int_{-\infty}^\infty du\, \rho(\xi,u)|v^{\text{eff}}(\xi,u)|(\partial_h^2 m^{\text{dr}}(\xi,u))^2, \tag{24}$$

where functions appearing in the integrand can be determined from the TBA integral equations. We show, later on, that the resulting rescaled TBA equations pertain to a particular regime of low-energy soliton modes of the isotropic Landau–Lifshitz ferromagnet.

### A. TBA functions of the Heisenberg chain at infinite temperature

The expression of the infinite-temperature TBA functions

$$\rho_s(\theta), \quad n_s(\theta) = \frac{1}{1+\eta_s(\theta)}, \quad v_s^{\text{eff}}(\theta) = \frac{\varepsilon'_s(\theta)}{2\pi\rho_s^{\text{tot}}(\theta)}, \quad m_s^{\text{dr}}(\theta), \tag{25}$$

for the isotropic Heisenberg spin-1/2 chain in an external magnetic field $-2hT\sum_i \hat{S}_i^z$ with $h > 0$ are reported, for example, in Ref.[8]. Here we spell them all out for completeness:

$$\eta_s(\theta) = \frac{\sinh(hs+h)^2}{\sinh(h)^2} - 1, \tag{26}$$

$$\rho_s(\theta) = \frac{2}{\pi}\frac{\sinh^4 h}{\sinh(2h))\sinh(h(s+1))}\left(\frac{s}{(4\theta^2+s^2)\sinh(hs)} - \frac{(s+2)}{(4\theta^2+(s+2)^2)\sinh(h(s+2))}\right), \tag{27}$$

$$\rho_s^{\text{tot}}(\theta) = \frac{2}{\pi}\frac{\sinh(h)}{\sinh(2h)\sinh(h(s+1))}\left(\frac{s}{(4\theta^2+s^2)\sinh(hs)} - \frac{(s+2)}{(4\theta^2+(s+2)^2)\sinh(h(s+2))}\right), \tag{28}$$

$$\varepsilon'_s(\theta) = 16\theta\frac{\sinh(h)\sinh(h(s+1))}{\sinh(2h)}\left(\frac{s}{(4\theta^2+s^2)^2\sinh(hs)} - \frac{(s+2)}{(4\theta^2+(s+2)^2)^2\sinh(h(s+2))}\right), \tag{29}$$

$$m_s^{\text{dr}}(\theta) = \frac{\sinh(h(s+1))((s+1)\cosh(h(s+1)) - \coth(h)\sinh(h(s+1)))}{\sinh^2(h(s+1)) - \sinh^2(h)}. \tag{30}$$

## B. The regulator: converting the quasiparticle sum into an integral

In taking the $h \to 0$ limit, we convert the sums over quasiparticles of the form $\sum_{s \geq 1} f_s$ into integrals using the rescaling $s \to \xi/h$,

$$\sum_{s \geq 1} f_s = \sum_{\xi = sh = h}^{\infty} f_{\xi/h}. \tag{31}$$

For small $h$ we find, using the Euler-MacLaurin formula to leading order

$$\sum_{\xi = h}^{\infty} f_{\xi/h} = \frac{1}{h} \int_0^\infty d\xi f(\xi) - \frac{f(0)}{2} + \mathcal{O}(h) \equiv \frac{1}{h} \int_0^\infty [d\xi]_h f(\xi), \tag{32}$$

where we where we assumed $f(\infty) = 0$ and we defined the regularized integral

$$\int_0^\infty [d\xi]_h f(\xi) \equiv \int_0^\infty d\xi f(\xi) - \frac{h}{2} \lim_{\xi \to 0^+} f(\xi), \tag{33}$$

with

$$f(\xi) = \lim_{h \to 0} f_{\xi/h}. \tag{34}$$

The "regularization" (32) is important to ensure the consistency of the semiclassical TBA equations which we derive below. As we subsequently demonstrate, in the context of classical soliton scattering theory, it plays the role of a *quantum correction*.

## C. Semiclassical TBA equation

*Infinite temperature.* Let us first restrict our consideration to the infinite temperature Gibbs state. The TBA equation for the functions $\eta_s(\theta)$, related to the Fermi occupation functions via $n_s(\theta) = 1/(1 + \eta_s(\theta))$, reads

$$\log \eta_s(\theta) = 2sh + \sum_{s'} \int d\theta' T_{s,s'}(\theta - \theta') \log(1 + 1/\eta_{s'}(\theta')). \tag{35}$$

The expression for the scattering kernel $T_{s,s'}$ can be found in the main text. By rescaling rapidities $\theta = u/h$, $\theta' = v/h$, the integral kernel takes the form

$$T_{s,s'}((u-v)/h) = \sum_{a=|s-s'|+2}^{s+s'-2} h \frac{4ah}{\pi((ah)^2 + 4u^2)} = \int_{h|s-s'|=|\xi-\zeta|}^{\xi+\zeta} \frac{d\tilde{a}}{2} \frac{4}{\pi} \frac{\tilde{a}}{\tilde{a}^2 + 4u^2} = \frac{1}{\pi} \log \frac{4u^2 + (\xi+\zeta)^2}{4u^2 + (\xi-\zeta)^2}. \tag{36}$$

discarding $\mathcal{O}(h)$ corrections. Note that the increment of the sum over $a$ equal 2. We refer to the result as the scattering kernel of the giant quasiparticles

$$T_{\xi,\xi'}^{\text{giant}}(u) = \frac{1}{\pi} \log \frac{4u^2 + (\xi+\xi')^2}{4u^2 + (\xi-\xi')^2}. \tag{37}$$

We furthermore define the rescaled $\eta$-functions

$$\eta_{s=\xi/h}(\theta = u/h) \to \eta(\xi, u)/h^2. \tag{38}$$

As $h \to 0$, the TBA equations can be written in the form

$$\log \eta(\xi, u) - 2 \log h = 2\xi + \int_0^\infty d\xi' \int_{-\infty}^\infty dv\, T_{\xi,\xi'}^{\text{giant}}(v - v') \eta^{-1}(\xi', v) - \frac{h}{2} \lim_{\xi' \to 0^+} \int_{-\infty}^\infty dv\, T_{\xi,\xi'}^{\text{giant}}(v - v') \eta^{-1}(\xi', v). \tag{39}$$

Taking into account that $\eta(\xi, u)$ does not depend on $u$ in the limit of infinite temperature, we can perform the integral over $v$,

$$\int dv\, T_{\xi,\xi'}^{\text{giant}}(u - v) = \xi + \xi' - |\xi - \xi'| = 2\min[\xi, \xi'], \tag{40}$$

yielding a simple infinite-temperature TBA equation for $\eta(\xi)$ pertaining to the giant quasiparticles

$$\log \eta(\xi) - 2\log h = 2\xi + \int_0^\xi d\xi' \frac{2\xi'}{\eta(\xi')} + \int_\xi^\infty d\xi' \frac{2\xi}{\eta(\xi')} - \frac{h}{2} \lim_{\xi' \to 0^+} \frac{2\xi'}{\eta(\xi')}. \tag{41}$$

A similar equation has appeared in the context of the solitons (breathers) of the classical sine-Gordon equation[9]. Following this reference, we differentiate twice with respect to $\xi$ to obtain a differential equation

$$(\eta'/\eta)' = -2/\eta, \tag{42}$$

which is in turn solved by an ansatz $\eta(\xi) = \left(\frac{\sinh(A\xi+B)}{A}\right)^2$ for appropriate constants $A$ and $B$. By substituting this expression for $\eta(\xi)$ in eq. (61) we find

$$\int_0^\xi d\xi' \frac{2\xi'}{\eta(\xi')} + \int_\xi^\infty d\xi' \frac{2\xi}{\eta(\xi')} - \frac{h}{2} \lim_{\xi' \to 0^+} \frac{2\xi'}{\eta(\xi')} = -2A\xi - 2\log\sinh B + 2\log\sinh(A\xi + B) - \frac{h}{2} \lim_{\xi' \to 0^+} \frac{2\xi'}{\eta(\xi')}, \tag{43}$$

which implies $A = 1$, $B = \operatorname{arcsinh}(h)$. Therefore, we find, at infinite temperature and for $h \ll 1$, the following result

$$\eta(\xi) = \sinh^2(\xi + h). \tag{44}$$

This indeed agrees precisely with the scaling limit of the exact solution for $\eta_s(\theta)$ in the quantum spin chain at infinite temperature, given by eq. (26). Note that

$$\lim_{\xi \to 0^+} \frac{2\xi}{\eta(\xi)} = 0, \tag{45}$$

so there is no additional correction entering when converting the quasiparticle sum to an integral over $\xi'$ using the regularization prescription (32).

*Finite temperature.* The above procedure likewise applies at any temperature $T$, where the rescaled TBA equations take the form

$$\log \eta(\xi, u) - 2\log h = 2\xi - \frac{h}{T} e^{\text{giant}}(\xi, u) + \int_0^\infty d\xi' \int_{-\infty}^\infty dv\, T^{\text{giant}}_{\xi,\xi'}(v - v') \eta^{-1}(\xi', v), \tag{46}$$

with bare energy of giant quasiparticles

$$e^{\text{giant}}(\xi, u) = \frac{2\xi}{\xi^2 + 4u^2}. \tag{47}$$

The solution to this equation is now rapidity-dependent in general, and we expect temperature to enter the part that regulates the integrals at small $\xi'$. More specifically, we expect that for small $\xi'$, $\eta(\xi', v) \simeq \sinh^2(\xi' + he^{h/Tf(v)})$ which will produce a factor $\log h + h/Tf(v)$ upon integration over $\xi'$.

### D. Free energy and spin susceptibility of giant quasiparticles

A closed formula for the free energy of the quantum spin chain, valid in any local equilibrium state, reads

$$-f/T = \sum_{s \geq 1} \int \frac{d\theta}{2\pi} \frac{4s}{s^2 + 4\theta^2} \log(1 + 1/\eta_s(\theta)). \tag{48}$$

With the same rescaling of variables as above, we obtain the following contribution of the giant quasiparticles

$$-f_{\text{giant}}/T = \frac{2h}{\pi} \int_0^\infty [d\xi]_h \int_{-\infty}^\infty du \frac{\xi}{\xi^2 + 4u^2} \frac{1}{\eta(\xi, u)}. \tag{49}$$

At infinite temperature, where we have $\eta(\xi) = \sinh(\xi + h)^2$, we find

$$-f_{\text{giant}}/T = h \int_0^\infty [d\xi]_h \frac{1}{\eta(\xi)} = h \left( \coth h - 1 - \frac{h}{2} \frac{1}{\sinh^2 h} \right). \tag{50}$$

In the limit $h \to 0$ this gives $-f_{\text{giant}}/T = 1/2$, which is less than the full value $-f/T = \log 2$. This indicates that the giant quasiparticles do not fully saturate the value of the free energy, and there are also contributions from 'small' bound magnons with bare magnetization $s \sim \mathcal{O}(1)$, namely

$$f = f_{\text{small}} + f_{\text{giant}}. \tag{51}$$

Interestingly however, spin susceptibility at half filling comes solely from the latter contribution to free energy. We can indeed show that we have at infinite temperature and $h \to 0$

$$\chi \equiv \chi_{\text{giant}} = -1/(4T)\partial_h^2 f_{\text{giant}} = \frac{1}{4}, \tag{52}$$

This shows, in contrast to the free energy, that the contribution of small quasiparticles entirely vanishes as $h \to 0$.

### E. Semiclassical dressing equations for energy and momentum

Other types of integral (dressing) equations in the TBA formalism which admit analogous semiclassical forms. For example, the Bethe equation for the total density of states in the quantum chain reads

$$\rho_s^{\text{tot}}(\theta) = a_s(\theta) - \sum_{s'} \int d\theta' \, T_{s,s'}(\theta - \theta') n_{s'}(\theta') \rho_s^{\text{tot}}(\theta'), \tag{53}$$

with the source term now given by the rapidity derivative of the bare momentum $p_s(\theta)$,

$$a_s(\theta) \equiv \frac{1}{2\pi} \frac{\partial p_s(\theta)}{\partial \theta} = \frac{1}{2\pi} \frac{4s}{s^2 + 4\theta^2}. \tag{54}$$

Introducing the rescaled density of state,

$$\rho_{s=\xi/h}^{\text{tot}}(\theta = u/h) \to h^2 \rho^{\text{tot}}(\xi, u), \tag{55}$$

and using the proper regularization of the integral (32) over $\xi'$, we have

$$\rho^{\text{tot}}(\xi, u) = \frac{1}{h} \frac{1}{2\pi} \frac{4\xi}{\xi^2 + 4u^2} - \int_0^\infty [d\xi']_h \int_{-\infty}^\infty dv \, T_{\xi,\xi'}^{\text{giant}}(v - v') \eta(\xi', v)^{-1} \rho^{\text{tot}}(\xi', v). \tag{56}$$

This integral equation implies the following relation

$$\lim_{h \to 0} h \int_0^\infty [d\xi']_h \int dv \frac{1}{\pi} \log \frac{4(u-v)^2 + (\xi+\xi')^2}{4(u-v)^2 + (\xi-\xi')^2} \eta(\xi', v)^{-1} \rho^{\text{tot}}(\xi', v) = \frac{1}{2\pi} \frac{4\xi}{\xi^2 + 4u^2}. \tag{57}$$

To verify that this integral equation is consistent with the rescaled thermodynamic functions of the quantum chain, we have checked numerically and analytically that this equation is satisfied at infinite temperature by the rescaled quantities $\eta(\xi') = \sinh^2(\xi' + h)$ and $\rho^{\text{tot}}(\xi', v) = 2(-4v^2 + \xi'^2 + 4v^2\xi' \coth \xi' + \xi'^3 \coth \xi')/(\pi(4v^2 + \xi'^2)^2)$, where the latter has been obtained by expanding the exact expression (28) for $\rho_{\xi/h}^{\text{tot}}(u/h)$ at small $h$.

The derivative of the dressed energy, $\varepsilon'_s(\theta)$, satisfies an integral equation analogous to eq. (53) with the bare energy derivative as the source term, i.e. function $\pi a'_s(\theta)$. In the scaling regime, the latter behaves as $h^2$ as $h \to 0$. Putting $\varepsilon'_s(\theta) = h^3 \varepsilon'(\xi, u)$, we find

$$\lim_{h \to 0} h \int_0^\infty [d\xi']_h \int dv \frac{1}{\pi} \log \frac{4(u-v)^2 + (\xi+\xi')^2}{4(u-v)^2 + (\xi-\xi')^2} \eta(\xi', v)^{-1} \varepsilon'(\xi', v) = -\frac{16u\xi}{(4u^2 + \xi^2)^2}. \tag{58}$$

It is straightforward to verify, see eq. (29), that this equation is also satisfied by the infinite-temperature solution

$$\varepsilon'(\xi, v) = -16v \frac{-4v^2 + 3\xi^2 + 4v^2\xi \coth \xi + \xi^3 \coth \xi}{(4v^2 + \xi^2)^3}. \tag{59}$$

We re-emphasize the importance of keeping the parameter $h$ finite before carrying out integration, as it regulates the divergence of the integral over $\xi'$ stemming from the divergent factor $\eta^{-1}(\xi, u) \sim 1/\xi^2$. We stress moreover that the proper regularization of the integral in eq. (56) is crucial to guarantee consistency with the the rescaled thermodynamic functions of the quantum chain in the limit of infinite temperature.

## F. Semiclassical equation for the dressed magnetization

Finally, we spell out the rescaled equation for the dressed magnetization. This one is somewhat simpler as the source term is simply given by $s$ (bare magnetization), and now $m^{\mathrm{dr}}_{s/h}$ scales as $1/h$ for small $h$. At infinite temperature, we can integrate over the rapidity variable to find

$$m^{\mathrm{dr}}(\xi) = \xi - \int_0^\xi d\xi' \frac{2\xi'}{\eta(\xi')} m^{\mathrm{dr}}(\xi') - \int_\xi^\infty d\xi' \frac{2\xi}{\eta(\xi')} m^{\mathrm{dr}}(\xi'). \tag{60}$$

Differentiating twice and using that at infinite temperature $\eta(\xi) = \sinh^2(\xi + h)$, we arrive at the following differential equation

$$\partial_\xi^2 m^{\mathrm{dr}} = \frac{2}{\sinh^2 \xi} m^{\mathrm{dr}}, \tag{61}$$

whose solution reads $m^{\mathrm{dr}}(\xi) = \frac{\xi}{\tanh \xi} - 1$, in agreement with the result obtained by rescaling the analytical solution given by eq. (30).

## III. THERMODYNAMIC SOLITON GAS AND QUANTUM-CLASSICAL CORRESPONDENCE

This section is devoted to the continuous isotropic Landau-Lifshitz ferromagnet, which is an effective classical theory governing the semi-classical eigenstates in the Heisenberg spin-$1/2$ chain. The Hamiltonian of the Landau-Lifshitz model in an external magnetic field in the $z$-direction $\vec{B} = B\,\vec{e}_z$ reads

$$H_{\mathrm{LL}} = \frac{1}{2} \int dx\, \partial_x \vec{S}(x) \cdot \partial_x \vec{S}(x) + B \int dx\, (1 - S^z(x)), \qquad \vec{S} \cdot \vec{S} = 1. \tag{62}$$

The corresponding equation of motion

$$\partial_t \vec{S} = \{\vec{S}, H\} \equiv -\vec{S} \times \frac{\delta H_{\mathrm{LL}}}{\delta \vec{S}} = \vec{S} \times \partial_{xx} \vec{S} + \vec{S} \times \vec{B}. \tag{63}$$

is a paradigmatic example of a nonlinear completely integrable PDE. The inverse scattering method for its integration has been developed in 10 (see also 11).

In order to integrate eq. (63) and classify the types of solutions one has to additionally prescribe the boundary conditions for the spin field $\vec{S}(x, t)$. Typically one assumes non-compact space-time with $\vec{S}(x, t)$ decaying exponentially into the asymptotic ferromagnetic vacuum $\vec{S}_0$, $\lim_{|x|\to\infty} \vec{S}(x, t) = \vec{S}_0$. By virtue of the global $SO(3)$ symmetry of Hamiltonian (62), $\vec{S}_0$ may indeed point to an arbitrary direction on a 2-sphere. In a finite volume, with $\vec{S}(x,t)$ subjected to (quasi)periodic boundary conditions, the integration of eq. (63) requires tools of algebraic geometry.

In the non-compact case with decaying fields, the spin field $\vec{S}(x)$ can be decomposed in term of 'non-linear Fourier modes'. The spectrum comprises a discrete part of non-linear interacting waves called *solitons* and a continuous spectrum of radiative modes. The finite-volume solutions on the other hand take the form of certain hyperelliptic waves whose parameters correspond to moduli of (hyperelliptic) algebraic curves. The spectrum of nonlinear modes decomposes into classes of finite-gap solutions, corresponding to Riemann surfaces of finite genus. Strictly speaking, soliton mode cannot form in finite volume. They nonetheless emerge in the infinite-volume limit as a degenerate case of an elliptic wave.

The aim of the following consideration is to obtain an integral dressing equation describing a dense gas of soliton modes in local thermal equilibrium. Deriving hydrodynamic equation of motion for local equilibrium states from first principles is not an easy task, and can be achieved by taking a particular type of thermodynamic scaling limit of an 'nearly-degenerate high-genus spectral curve'[12]. As we shall see, the problem simplifies quite significantly if one retains only soliton modes.

### A. One-soliton solution

By assuming asymptotic decay to $S_0 = S^z$ at infinity $|x| \to \infty$, the simplest discrete spectrum comprises a single mode. Physically, this describes a one-soliton solution. The soliton profile (in the absence of an external field) takes

the form[10]

$$S^z(x,t) = 1 - \frac{A_z}{\cosh^2(\varphi(x,t))}, \qquad S^-(x,t) = \frac{A_- e^{-i\phi(x,t)}}{\cosh^2(\varphi(x,t))}\Big[ia\cosh(\varphi(x,t)) + b\sinh(\varphi(x,t))\Big], \tag{64}$$

with amplitudes

$$A_z = \frac{2b^2}{a^2+b^2}, \qquad A_- = \frac{2b}{a^2+b^2}, \tag{65}$$

and dynamical angles

$$\varphi(x,t) = \frac{1}{\Gamma}(x - vt + x_0), \qquad \phi(x,t) = ax + (a^2 - b^2)t + \varphi_0. \tag{66}$$

The velocity and width of a soliton read

$$v = 2a, \qquad \Gamma = \frac{1}{b}, \tag{67}$$

respectively. The one-soliton solution (64) is two-parameter family of solutions to eq. (63) characterized by quantities $a, b \in \mathbb{R}$. These can be understood as *action variables*, whereas the associated linearly-evolving phases $\phi(x,t)$ and $\varphi(x,t)$ are *angle variables*. Combining $a$ and $b$ into a single complex action we introduce

$$\tau = \frac{1}{a - ib} \equiv \lambda + i\mu, \qquad \lambda \in \mathbb{R}, \quad \mu \in \mathbb{R}_+. \tag{68}$$

In the new parametrization

$$A_z = \frac{2\mu^2}{\lambda^2 + \mu^2}, \qquad A_- = 2\mu, \qquad v = \frac{2\lambda}{\lambda^2 + \mu^2}, \qquad \Gamma = \frac{\lambda^2 + \mu^2}{\mu}. \tag{69}$$

The phase-space representation of the bare energy, momentum and magnetization (measured with respect to vacuum value $S^z = 1$) read

$$E = \frac{1}{2}\int dx\, \partial_x \vec{S}(x) \cdot \partial_x \vec{S}(x), \qquad P = \int dx\, \frac{S^x \partial_x S^y - S^y \partial_x S^x}{1 + S^z}, \qquad Q^z = \int dx\,(1 - S^z(x)). \tag{70}$$

The charges carried by a single soliton (cf. eq. (64)) are therefore

$$e = e(\lambda, \mu) = \frac{4\mu}{\lambda^2 + \mu^2}, \qquad p = p(\lambda, \mu) = 4\arctan(\mu/\lambda), \qquad q^z = 4\mu, \tag{71}$$

respectively. This yields the following dispersion relation

$$e(\lambda, \mu) = \frac{1}{\mu^2}\sin^2(p/4), \tag{72}$$

and the bare velocity of propagation is

$$v = \frac{\partial e_\mu(p)}{\partial p} = \frac{1}{\mu}\sin(p/2) = \frac{2\lambda}{\lambda^2 + \mu^2}. \tag{73}$$

With the addition of the magnetic field in the $z$-direction of magnitude $B = h_{\rm cl}\, T_{\rm cl}$ the dispersion acquires a shift

$$e_\mu(\lambda; h) = e(\lambda, \mu) - h_{\rm cl}\, T_{\rm cl}\, q^z = \frac{4\mu}{\lambda^2 + \mu^2} - 4h_{\rm cl}\, T_{\rm cl}\, \mu. \tag{74}$$

## B. Scattering phase shift

Considering only the multi-soliton spectrum, one can introduce the notion of the many-body scattering matrix. Indeed, owing to the underlying integrability, the $n$-particle $\mathcal{S}$-matrix completely *factorizes* into a sequence of 2-particle collisions,

$$\mathcal{S}(\tau_1,\ldots,\tau_n) = \prod_{1\leq j,k\leq n} \mathcal{S}^{(2)}(\tau_j,\tau_k), \tag{75}$$

where we have introduced the two-particle $\mathcal{S}$-matrix,

$$\mathcal{S}^{(2)}(\tau,\tau') = e^{i\Phi(\tau,\tau')}. \tag{76}$$

Here $\Phi$ is referred as the scattering phase, which is a function of action variables $\tau = \lambda + i\mu$ and $\tau' = \lambda' + i\mu'$ of the incident solitons. A convenient operational prescription of the scattering phase is via the time-delay that the 'first' soliton experiences after collision with the 'second' soliton[13]

$$\delta t(e,e') = \frac{\partial \Phi(e,e')}{\partial e}. \tag{77}$$

Introducing the (manifestly positive) *scattering length*

$$d(\tau,\tau') = \left|\frac{\partial \Phi(\tau,\tau')}{\partial p(\tau)}\right| = \frac{\partial_\lambda \Phi(\tau,\tau')}{|\partial_\lambda p(\tau)|}, \tag{78}$$

the asymptotic position displacement $\delta x$ attributed to each collision is given by

$$\delta x(\tau,\tau') = \mathrm{sign}(v(\tau)-v(\tau'))d(\tau,\tau'), \tag{79}$$

where, for $v > v'$, we have

$$\delta x(\tau,\tau') = \Gamma(\tau)\,\delta\varphi(\tau,\tau'), \qquad \delta\varphi(\tau,\tau') = \log\left|\frac{\tau-\bar{\tau}'}{\tau-\tau'}\right|^2. \tag{80}$$

This readily implies

$$\partial_\lambda \Phi(\tau,\tau') = 4\,\delta\phi(\tau,\tau') = 4\log\left[\frac{(\lambda-\lambda')^2+(\mu+\mu')^2}{(\lambda-\lambda')^2+(\mu-\mu')^2}\right]. \tag{81}$$

## C. Thermodynamic soliton gas

Owing to the fact that the multi-soliton dynamics realizes a completely elastic scattering theory with a fully factorizable $\mathcal{S}$-matrix[14], one can make use of the Yang–Yang's path-integral approach to thermodynamics[15]. The logic parallels that of integrable quantum models such as Lieb–Liniger Bose gas of Heisenberg spin chain[16,17]: a *thermodynamic state* is understood as a finite-density ensemble of solitons, describing $\mathcal{O}(\ell)$ excited soliton modes in a a thermodynamically large system $\ell \to \infty$, with the condition that the averages of all local conservation laws per unit length are kept constant. Notice that at the level of thermal averages any dependence on initial conditions (i.e. values of the angle variables) entirely drops out.

The thermodynamic partition sum can be conveniently evaluated with the standard saddle-point technique by first expressing it in terms of soliton action variables and subsequently switching from a microscopic description in terms of discrete actions to a coarse-grained description in terms of their spectral densities. Introducing the Boltzmann factor $\eta = \exp(\varepsilon)$, where $\varepsilon$ is referred to as the dressed energy, and minimizing the Shannon entropy, yields the following Fredholm integral equation

$$\log\eta(\lambda,\mu) = h_{\mathrm{cl}}\,q^z - \frac{e(\lambda,\mu)}{T_{\mathrm{cl}}} + \int_0^\infty d\mu' \int_{-\infty}^\infty d\lambda' T^{\mathrm{cl}}_{\mu,\mu'}(\lambda,\lambda')[\eta(\lambda',\mu')]^{-1}, \tag{82}$$

where $T_{\mathrm{cl}}$ is temperature, $h_{\mathrm{cl}}$ plays the role of chemical potential, while the integral kernel

$$T^{\mathrm{cl}}_{\mu,\mu'}(\lambda,\lambda') = \frac{\partial_\lambda \Phi(\tau,\tau')}{2\pi}, \tag{83}$$

is given by the differential solitons scattering shift.

## D. Soft solitons

We now analyze eqs. (82) in the vicinity of a non-magnetized sector, that is in the limit $h \to 0$. Our aim is to retrieve the dressing equations for a finite-density thermal gas of giant quasiparticles obtained in the previous section. This will require us to consider a particular low-energy regime governed by wide and shallow solitons. This can be achieved by taking two scaling limits $h_{\rm cl} \to 0$ and $\lambda, \mu \to \infty$ simultaneously, while keeping the ratios

$$u = \lambda\, h_{\rm cl}, \qquad \xi = h_{\rm cl}\, q^z/2 = 2\mu\, h_{\rm cl}, \tag{84}$$

finite. This will project out the low-energetic solitons with $e \sim h_{\rm cl}$, large width $\Gamma \sim 1/h_{\rm cl}$, and large magnetization $q^z \sim 1/h_{\rm cl}$. The bare momentum $p$ on the other hand stays finite, and the scattering phase shift remains unaltered. At the same time we need to rescale $\eta(\lambda, \mu) \to \eta(\xi, u)/h_{\rm cl}^2$ to compensate for the change of variables. The appropriately rescaled version of eq. (82) then has the form

$$\log \eta(\xi, u) = 2\log h_{\rm cl} + 2\xi - \frac{h_{\rm cl}}{T_{\rm cl}} e^{\rm cl}(\xi, u) + \int_0^\infty d\xi' \int_{-\infty}^\infty du'\, T^{\rm cl}_{\xi,\xi'}(u, u')[\eta(\xi', u')]^{-1}, \tag{85}$$

with

$$e^{\rm cl}(\xi, u) = \frac{8\xi}{4u^2 + \xi^2}, \qquad T^{\rm cl}_{\xi,\xi'}(u, u') = \frac{1}{\pi}\log\left[\frac{4(u-u')^2 + (\xi + \xi')^2}{4(u-u')^2 + (\xi - \xi')^2}\right], \tag{86}$$

being the bare energy of soft solitons and their scattering kernel, respectively.

Let us compute the contribution of soft solitons to the Gibbs free energy density,

$$-f_{\rm soft}/T_{\rm cl} = \frac{h_{\rm cl}}{\pi}\int_0^\infty d\xi \int_{-\infty}^\infty du\, \frac{\xi}{4u^2 + \xi^2}\, \frac{1}{\eta(\xi, u)}. \tag{87}$$

Here the integral over $\xi$ is an ordinary integral, in contrast to the quantum case (49). Using these expressions, we deduce the static spin susceptibility in the infinite-temperature limit,

$$\lim_{h_{\rm cl}\to 0} \chi(h_{\rm cl}) = -\partial^2_{h_{\rm cl}}(f_{\rm soft}/T)\Big|_{h_{\rm cl}=0} = \frac{1}{2}\partial^2_{h_{\rm cl}}\big(h_{\rm cl}(\coth h_{\rm cl} - 1)\big) = \frac{1}{3}. \tag{88}$$

Soft solitons exhaust the full spin susceptibility in the $h \to 0$ limit. Crucially however, this results differs from static spin susceptibility of the quantum spin chain (as it, of course, should); the difference between can be attributed to the regularization of the $\xi$-integration around $\xi = 0$.

We notice that this gas of soft solitons coincides precisely with the giant quasiparticles we have identified in the quantum Heisenberg chain, as they have identical scattering kernel and dispersion relation. The momentum and energy differ between the quantum and classical cases by numerical multiplicative factors, which can be absorbed by a redefinition of units of space and time (or energy). We do not fully understand the origin of these numerical factors, and it would be interesting to clarify this point in future works.



\end{document}